\documentstyle[aps,12pt,epsfig]{revtex}

\newcommand{\beq}{\begin{equation}}
\newcommand{\eeq}[1]{\label{#1} \end{equation}}
\newcommand{\beqar}{\begin{eqnarray}}
\newcommand{\eeqar}[1]{\label{#1} \end{eqnarray}}

\begin{document}
\normalsize



\title{The Mini-Jet Scale  and Energy Loss at RHIC\\
 in the HIJING and RQMD Models} 
\author{~V. Topor Pop$^{1}$,  M. Gyulassy$^2$, 
J. Barrette$^{1}$, C. Gale$^{1}$,\\ 
~X. N. Wang$^3$, N. Xu$^3$, K. Filimonov$^3$}
\address{
$^1$McGill, University, Montreal, Canada, H3A 2T8\\
$^2$Physics Department,
Columbia University, New York, N.Y. 10027\\
$^3$Nuclear Science Division, LBNL, Berkeley, CA 94720\\
}
\date{August 15, 2003}

\maketitle

\vskip 0.3cm
\begin{abstract}
Recent data from Relativistic Heavy Ion Collider (RHIC)
 on rapidity and transverse momentum distributions
of hadrons produced in ultra-relativistic reaction of both Au+Au
and p+p are compared to predictions of the HIJING
and RQMD models. 
The original default mini-jet scale $p_0=2$ GeV/c 
and energy loss, $dE/dx=2$ GeV/fm in HIJING lead to a too rapid
growth of the multiplicity with energy. RQMD model without mini-jet
leads on the other hand to a too slowly increasing multiplicity with energy.
Therefore, we study what variations of $p_0$ and $dE/dx$
are required in HIJING to
account for the observed $N_{part}$ and $\sqrt{s}$ dependence of
the global $dN_{ch}/dy$ and $dE_T/dy$ observables
as well as the jet quenching pattern out to $p_T\sim 8$ GeV.
We show that a slight increase of 
 $p_{0}$ from the default 2.0 GeV/c at $\sqrt{s}=130$A GeV
 to 2.2 GeV/c at $\sqrt{s}=200$A GeV
is sufficient to account for the bulk observables. Jet quenching 
of $\pi^0$ above $p_T>4$ GeV/c at 200A GeV is found to be well reproduced 
in central $Au+Au$ by  the
original HIJING default prediction.  However, the 
effective surface emission in the HIJING model formulation of energy loss
over-predicts the quenching in more peripheral collisions.
Neither HIJING nor HIJING/B\=B
can account  for the observed anomalous excess of moderate 
$p_T<4$ GeV/c baryons.
The agreement of RQMD with data in this $p_T$ region is shown to be fortuitous 
because, without mini-jets, it fails to reproduce the $p+p$ spectrum.

\end{abstract}


\section{Introduction}

Qualitatively new phenomena 
 have been discovered\cite{qm01,qm02} in ultra-relativistic 
nuclear collisions at
total center of mass energy $E_{cm}=130A-200$A GeV at the Relativistic
Heavy Ion Collider (RHIC) and interpreted as evidence
for the formation of ultra-dense QCD matter. 
The extensive systematics of both  bulk global observables
and high transverse momentum phenomena were reported by  
PHENIX\cite{qm2001_za,phenix01_20,phenix01_30,phenix01_40,Adcox:2002ms,phe02_raa_1,bazilev03,jia03,ad03_pi0_1,ad03_pi0_2,jia03_1}, 
STAR\cite{qm2001_ha,Ackermann:2000tr,star_ad,star01_raa,star_dun,qm02_kunde,jet_star03,Adler:2002ct,Adams:2003kv,star_dAu}, 
PHOBOS\cite{qm2001_ro,back_00,back00_20,phobos_ge,back01_30,Back:2002gz,Back:2003qr,phobos_dAu} 
and
BRAHMS\cite{qm2001_vi,brahms12_01,jham_ch03,beard_pbar03,brahms_03}.
RHIC data are now available not only for $Au+Au$ collisions 
but also for $p+p$  with high statistics, to accurately calibrate the
magnitude of nuclear effects, and also $d+Au$
\cite{jia03_1,star_dAu,phobos_dAu,brahms_03} 
needed as a control experiment to
separate initial versus final state interaction effects.

In this paper we compare  predictions of the 
HIJING\cite{hij92_99} and 
RQMD\cite{sorge95} models to data. These nuclear collision event generators 
were developed a decade ago and continue to be 
useful tools for detector design and interpretation of experimental results
because they simulate complete exclusive event characteristics 
at SPS and RHIC.
HIJING, incorporating pQCD mini-jet production,
 is constrained to reproduce essential features of 
$p+p$ data over a wide energy range.
RQMD is designed to simulate only soft multi-particle 
production at lower energies,
but is of interest since it incorporates a  model of final state interaction. 
No attempt is made here to review the many other models
developed since the release of HIJING and RQMD. We refer the reader to Refs. 
\cite{bass99_1,miklos03,heinz03,munz03,iancu03,xin_th02,vitev00,Gyulassy:2000er,Gyulassy:2001nm,Vitev:2001zn,Gyulassy:2001kr,Vitev:2002pf,Vitev:2003xu,xin_th01,xin_th03,xin_li02,Wang:2002ri,Levai:2003at,ekrt01,eskola_th01,dima_th01,dima_th02,dima_th03,dima_th04,dima_th05,cassing99_1,lin01,Lin:2001zk,Lin:2002gc,Lin:2001yd,Molnar:2001ux,Molnar:2002bz,bass98,bleicher99_2,armesto00_1,capella_01,kapusta2000_1,kahana2000} for a broader perspective.
 
Our goal is to test specific predictions within  HIJING and RQMD models, 
made well in advance of the data.  
The predictions  involved estimating key physical
parameters controlling the dynamics. An important aim of this
paper is to investigate what 
adjustments of those parameters may be required in light of the new data.
The  two main parameters of HIJING 
that we concentrate on in this paper are (1) the separation scale,
 $p_0$ between the perturbative (pQCD)  mini-jet processes and 
the phenomenological soft (beam jet fragmentation) processes and (2) 
the energy loss, $dE/dx$, of high transverse momentum
partons propagating through the dense (quark-gluon plasma) 
medium produced in the reaction.

The mini-jet separation scale, $p_0\approx 2$ GeV/c, 
assumed in HIJING\cite{hij92_99} controls the
$\sqrt{s}$ dependence of the bulk multiplicity and 
transverse energy observables
as well as their centrality dependence on the number of 
wounded nucleon participants
$N_{part}$. RQMD\cite{sorge95} and its UrQMD\cite{bass99_1,bass98} extensions
 assume in effect  
that $p_0\rightarrow \infty$ and therefore neglect the power law tails 
due to pQCD mini-jets. 
The default constant value of $p_0=2$ GeV/c in HIJING was fixed 
by fitting $pp$ data through Tevatron energies\cite{hij92_99}, and 
it was assumed to be {\em independent}
of both $\sqrt{s}$ and $A$ in order to 
predict multiparticle observables in $AA$ that extrapolate 
down accurately to $A=1$ 
to reproduce experimentally known $pp$ data.  In RQMD, there was no attempt
to fit collider energy pp data. Our approach differs 
in this major respect
from many recent models that fail to account for multi particle phenomena
in $p+p$ collisions. Our philosophy is that due to the myriad of 
dynamical complexities already 
displayed in ''elementary'' $p+p$ reactions,
 any model proposed to explain $A+A$ collisions 
must extrapolate accurately down to $A=1$. 
This is because one of the very few experimental 
knobs in heavy ion reactions
is the variation of the impact parameter through the participant number
dependence of observables. In peripheral collisions 
the observables necessarily approach
their value in $p+p$ collisions.

Recently, several models were developed that challenge 
the assumption that the separation scale $p_0$ is independent  of both $\sqrt{s}$ and $A$ as in HIJING,
Refs.\cite{ekrt01,eskola_th01} generalized
the mini-jet scale by allowing it to
vary dynamically  by introducing the  hypothesis of 
{\em final state saturation} (FSS) of the produced mini-jet density 
per unit area.
That saturation scale was predicted to be $p_s(s,A)=1.1 {\rm \; GeV}\;
(\sqrt{s}/200)^{0.128}(A/200)^{0.191}$.
This hypothesis leads to a specific prediction
for the initial parton distribution in $A+B$ collisions.
With the additional hypothesis that this initial condition
evolves according to local thermal  hydrodynamics, 
 the $p_T$ integrated global observables in central $Au+Au$ were shown to be well account for.
However, the breakdown of hydrodynamics and the smallness
of $p_s$ ($p_s< 1$ GeV/c) in peripheral collisions prevents the model
from predicting correctly peripheral $A+A$ collisions and $p+p$. 

An alternate {\em initial state saturation} (ISS) hypothesis 
for the variation of the separation scale was introduced in
Refs.\cite{dima_th01,dima_th02,dima_th03,dima_th04,dima_th05,lmc_v94,yuri96}.
In this picture, $p_0$ is replaced by a gluon saturation scale
$Q_s(s,A)\approx 1.4 \; {\rm GeV} \; (\sqrt{s}/200)^{0.3} (A/200)^{1/6} $.
This saturation scale is 
considered to be the boundary of the classical Yang-Mills field domain.
Instead of hydrodynamics,  local parton hadron duality is assumed to
predict low $p_T\sim Q_s$ integrated bulk global observables. 
The  normalization of $Q_s$ was fixed by fitting the observed 
central 130A GeV Au+Au rapidity density. 
ISS was found to be more successful in describing the lower 
participant number and
$\sqrt{s}$ dependence of the rapidity density at RHIC than the FSS model
due to a particular low $Q^2\sim 1-2$ GeV$^2$  
dependence of the gluon structure function and fine structure coupling.
While preliminary extensions\cite{dima_th04} of this 
ISS model to the $p_T>Q_s$ regime could fit central 130A GeV $Au+Au$
without final state interactions,
the most recent data  on $d+Au$ reactions
\cite{jia03_1,star_dAu,phobos_dAu,brahms_03}
rule out a particular extension\cite{dima_th05} of the 
ISS model at mid rapidity in the range $2< p_T <10$ GeV/c.

 While neither ISS or FSS saturation models  can describe 
simultaneously the  global low $p_T$ and the 
hard high $p_T$ observables in both $p+p$ and $A+A$ collisions, 
they both provide strong motivation to test the effects of variations
of $p_0$ with both $s$ and $A$ in HIJING.
With this motivation, 
the first part of this paper will be to investigate, within the HIJING
model,  the effect  of relaxing the  $(s,A)$ independence of
the default constant $p_0=2$ GeV/c assumption. Importantly, 
we seek to do this while retaining our philosophy  that
critical features of multiparticle production in $p+p$ 
should be accounted for simultaneously in the same
model. Our first conclusion will be that the global 
130-200A GeV $Au+Au$ data
can be well accounted for by allowing a rather modest 
 10\% (A independent) enhancement of $p_0$ from $2$ 
GeV/c at 130A GeV to about $2.2$ GeV/c at 200A GeV.

The second part of this paper focuses on jet
quenching\cite{gyu90}. Jet quenching is one of 
the most striking 
new phenomena\cite{phenix01_40,ad03_pi0_1,star01_raa,jet_star03,Adams:2003kv,Back:2003qr}
discovered at RHIC. This effect was not  observed previously at 
lower energies. In fact, at SPS a strong (Cronin) enhancement 
of high $p_T$ $\pi^0$ was observed\cite{Aggarwal:2001gn}.

The first quantitative predictions of jet quenching with  HIJING
\cite{hij92_99,wang92} assumed a default
constant $dE/dx=2\;(1)$ GeV/fm gluon (quark) energy loss 
and was implemented by a simple string flip algorithm assuming a
mean free path of $\lambda=1$ fm.  
As recently reviewed in \cite{miklos03,baier95}, there has been considerable
progress since that time in computing the dependence 
of medium induced radiative energy loss,
$\Delta E(p_T,L, \rho_0)$, in
QCD\cite{vitev00,Gyulassy:2000er,Gyulassy:2001nm,xin_th01,xin_th03,baier95}.
The observable consequences 
of the dependence of $\Delta E$
on the jet energy, its propagation length, and the evolving parton density 
have been explored in pQCD models in which the plasma is 
assumed rather than calculated
dynamically. While the jet quenching algorithm in HIJING is
much more schematic, the model provides a useful theoretical
laboratory to study its observable consequences 
in the dynamical medium that it creates.
In the second part of this paper, we therefore explore what modifications 
of the default HIJING
assumptions are required in light of the new data.

As we show below, the default constant energy loss in HIJING 
accounts remarkably well
for the high $p_T$ $\pi^0$  suppression pattern in central $Au+Au$ at
200A GeV.
In fact, HIJING also accounts\cite{Gyulassy:1998nc} 
for the enhancement observed\cite{Aggarwal:2001gn}
at SPS that is otherwise puzzling according 
to pQCD estimates\cite{Wang:1998hs}.
However, we show below that HIJING fails to account for
the observed centrality dependence 
as well as for the anomalous baryon enhancement
observed up to 5 GeV. HIJING/B\cite{svance98,svance99,svance_cu}, 
with its implementation of baryon junctions\cite{rossi80,kharzeev96}, 
was tested to see if this mechanism could account 
for the baryon anomaly (see also \cite{Vitev:2001zn}).
However, the present version (HIJING/B\=Bv1.10) fails to account 
for the large transverse slopes of anti-baryons and does not reproduce
the ``baryon lump'' at moderate $p_T$ in the nuclear modification factor.

While RQMD\cite{sorge95} does not contain mini-jets, 
we investigate its predictions
because  it is one of first models to simulate 
final state  transport dynamics  of pre-hadrons and  hadrons
without assuming local equilibrium as in hydrodynamics.
It  was able to reproduce the directed and elliptic collective
flow systematics observed in Pb+Pb at SPS (17A GeV). However, 
as we show below,
the absence of hard pQCD processes leads to a much 
too weak beam energy dependence
at RHIC energies. It therefore also fails to account for the 
power law tails of the $p+p$ spectra 
at 200 GeV. We refer to Refs.  \cite{bass98,bleicher99_2} for a review
of application of RQMD and URQMD 
applied to reactions at AGS and SPS  energies. 

While not addressed directly in this paper, we call attention to the recent 
AMPT transport model\cite{lin01} that
incorporates mini-jet production and extends HIJING by 
including both parton cascading
and hadronic final state interactions. AMPT is under extensive  development 
and has been tested on a number of important 
RHIC observables\cite{lin01,Lin:2001zk,Lin:2002gc,Lin:2001yd}. 
However, problems with covariance of numerical solutions
involving ultra-relativistic  
parton cascading\cite{Molnar:2001ux,Molnar:2002bz}
require very high parton subdivision techniques which are unfortunately
beyond present computer power to solve with AMPT.

In order to bypass current  technical difficulties 
of predicting bulk collective phenomena via transport theory,
hydrodynamic models have also been extensively applied 
\cite{heinz03,landau53,bjorken83,stocker86,Kolb:2000fh,Huovinen:2001cy,Teaney:2001av,Huovinen:2003fa,Hirano:2003pw}.
The central simplifying dynamical assumption is that 
perfect local equilibrium is established and maintained
throughout the reaction. 
Therefore 
non-viscous hydrodynamics together with a Cooper-Frye statistical freeze-out
prescription \cite{frye74} are used to compute the expansion,
hadronization, and subsequent expansion until freeze-out. 
No attempt is made in such models to compute the initial
condition, but rather the initial entropy and baryon density are 
fit to the measured  rapidity distributions.
While such models cannot  predict
beam energy dependence of observables, they do predict 
striking collective flow
phenomena and their dependence on the QCD equation of state
(for a recent review of hydrodynamics at RHIC see Ref.\cite{heinz03}).
The first attempt at a hybrid combination hydrodynamics and jet quenching
was proposed in \cite{Gyulassy:2001kr}. Recently an important step forward 
is the development of a consistent 3+1D hydrodynamical
approach including QCD jet quenching \cite{Hirano:2003pw}.

Unlike hydrodynamics\cite{heinz03,Hirano:2003pw} 
or parton transport models\cite{Lin:2001zk,Molnar:2001ux}, 
neither HIJING nor RQMD can predict  the large amplitude elliptic flow
observed \cite{Ackermann:2000tr,Adler:2002ct,Adcox:2002ms,Back:2002gz}
at RHIC. 
Elliptic flow is especially sensitive to
early partonic final state interactions 
beyond the capability of these models.
However, azimuthally integrated inclusive spectra
are still interesting and can be addressed by models considered here.
We use the following versions of 
HIJING v1.37 \cite{hij92_99}, HIJING/B\=B v1.10 \cite{svance98,svance99} 
and RQMD v2.4 \cite{sorge95} for the computations reported below.

\section{Charged Particles Distributions and the Mini-Jet Scale}

Recent measurements of the rapidity density of 
charged particles in Au+Au collisions 
over the range of total nucleon-nucleon center of mass (c.m.)
energy $\sqrt{s_{NN}}$=56 GeV - 200 GeV, have been reported
\cite{phenix01_20,phenix01_30,phenix01_40,phe02_raa_1,star_ad,star01_raa,back00_20,phobos_ge,back01_30,brahms12_01}. 
Within the errors, an approximatively logarithmic rise of 
charged particle rapidity density 
{\em per participating baryon pair} 
($dN_{ch}/d\eta/0.5N_{part}$) with $\sqrt{s_{NN}}\,\,$ is observed over 
the full range of collision energies \cite{back01_30}.
In Refs.\cite{mik2000_1,mik_th01} 
the centrality dependence of this observable was proposed as a test of
 the nuclear enhancement of the mini-jet 
component as well as whether gluon saturation is reached at RHIC energies.
The predictions of different models varied prior to the data greatly from 
$dN_{ch}/dy \sim 700 - 1500$ at mid-rapidity for Au+Au central collisions
\cite{bass99_1}.

\begin{figure} [hbt!]
\vskip 0.5cm
\centering
\epsfig{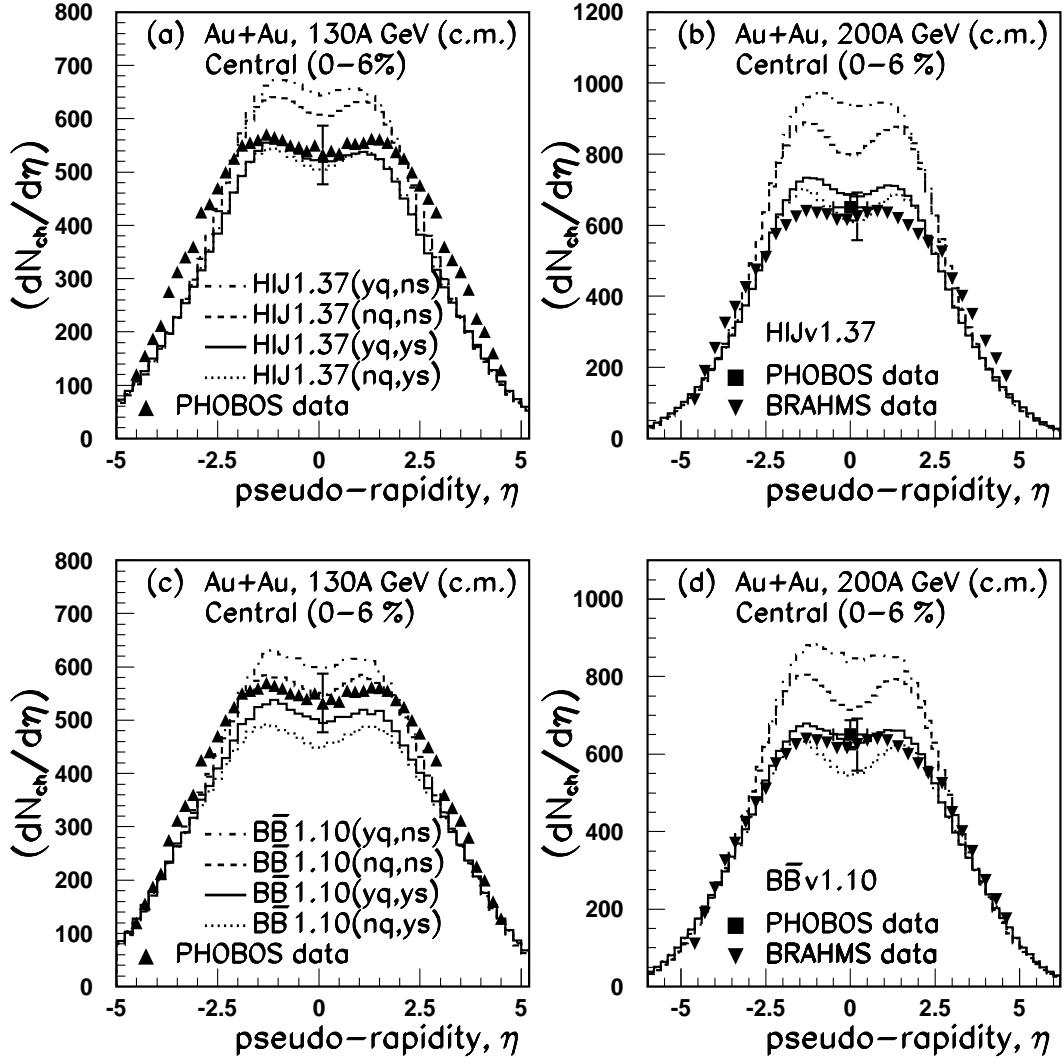} 
\vskip 0.5cm
\caption[rapidity distributions of all ch]{\small  
Charged hadron rapidity distributions for central (0-6\%) 
Au+Au collisions as function of c.m. energies.
The histograms show the theoretical predictions from
HIJING v1.37 (upper part) and HIJING/B\=B v1.10 (lower part) 
with (y) or without (n) effects of quenching (q)
or/and shadowing (s) included.
The data are from PHOBOS collaboration \cite{back00_20} (a), 
\cite{phobos_ge} (b) ,\cite{back01_30} (c) and BRAHMS
Collaboration \cite{brahms12_01} (c). 
The error bars at midrapidity include systematic uncertainties.
The others error bars of the order of 10-15 \% have been 
omitted for clarity.}
\label{fig:fig1_qm02}
\end{figure}

The predictions of HIJING (Figs 1a,b) and HIJING/B\=B (Figs 1c,d) 
with (y) or without (n) effects of quenching (q) or/and 
shadowing (s) are presented in Fig.~\ref{fig:fig1_qm02}.
The data from  PHOBOS \cite{back00_20,phobos_ge,back01_30}
and BRAHMS  \cite{brahms12_01} experiments 
at $\sqrt{s_{NN}}$=130 GeV and $\sqrt{s_{NN}}$=200 GeV
are shown  for comparison. In all cases,  
both quenching and possible parton shadowing 
influence the predicted pseudo-rapidity distribution, $dN_{ch}/d\eta$,
in this energy range. 
However, these effects work in 
opposite direction and thus partially cancel each other. 
Without shadowing as assumed in the default HIJING model,
 the flux of mini-jets with $p_T>2$ GeV/c is too high
and $dN_{ch}/d\eta$ is overestimated. Even with  the larger shadowing 
at the smaller x at  $\sqrt{s_{NN}}$=200 GeV, the 
gluon density enhancement resulting from mini-jet energy loss
leads to a 10-20\% overestimate of  the charged particle rapidity density
in going from 130A to 200A GeV in Fig. 1b. 
A similar tendency is seen in the B\=B version
of HIJING which, however, better accounts for the width of the 
rapidity distributions.
The width is sensitive to the nuclear fragmentation physics,
especially  baryon number transport from the beam
rapidities. HIJING/B\=B can better account for nuclear fragmentation
 by introducing the greater baryon stopping
power through the  baryon junction mechanism.

These data are also consistent with the 
initial state saturation ISS model \cite{brahms12_01,dima_th01,dima_th02}. 
However, the EKRT final state saturation model \cite{ekrt01} 
tends to over-predict the width of the rapidity distribution.

\begin{figure} [hbt!]
\vskip 0.5cm
\centering 
\epsfig{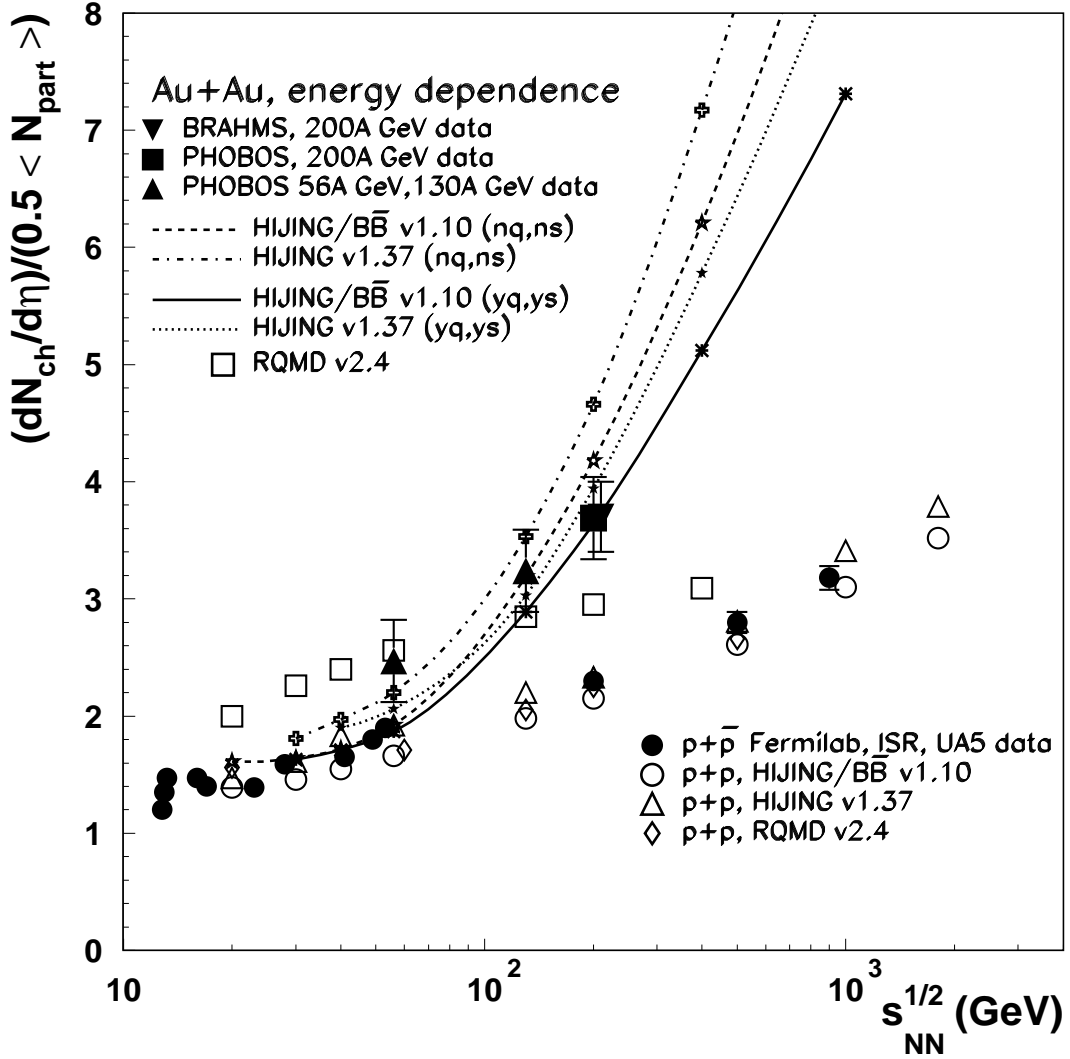} 
\vskip 0.5cm
 \caption[phobos dn ch comparison]{\small
Charged particle rapidity density 
{\em per participating baryon pair} versus the c.m. energy.
The predictions of HIJING/B\=B v1.10 {\em (yq,ys)}-full line,
HIJING/B\=B v1.10 {\em (nq,ns)}-dashed line,
HIJING v1.37 {\em (yq,ys)}-dotted line,
HIJING v1.37 {\em (nq,ns)}-dot dashed line, 
 and RQMD v2.4 ({\em open squares})  
are compared to data. 
The data for the central ($0-6\%$) 
Au+Au collisions, are from PHOBOS \cite{back00_20},\cite{back01_30},
and from BRAHMS \cite{brahms12_01};
the error bars include systematic uncertainties.
pp and p\=p data are from  ref. \cite{ua5,thome77,whit74,ua1};
the error bars are statistical only.} 
\label{fig:fig2_qm02}
\end{figure}

The energy dependence of the particle multiplicity is more easily seen in
Fig.~\ref{fig:fig2_qm02} where the central rapidity
density per participant pair vs $\sqrt{s}$ of both HIJING and RQMD
are compared to data. 
The PHOBOS data for the central (0-6\%) Au+Au collisions are from  
\cite{back00_20,phobos_ge,back01_30}.
The data for pp and p\=p are from \cite{ua5,thome77,whit74,ua1}.
 Beginning about $\sqrt{s_{NN}}$=100 GeV, the central Au+Au
collisions show a significantly larger particle density per
participant pair than in inelastic p\=p collisions.
The energy dependence predicted by
HIJING is strikingly different than that predicted by RQMD.
While RQMD predicts a very small increase 
over the range $\sqrt{s_{NN}}$=56 GeV - 200 GeV, 
HIJING predicts an increase of more 
than a factor of 1.3, which continue up to the highest energy calculated.
This increase in HIJING is due to copious mini-jets production in
$A+A$ collisions.
RQMD fails to describe the trend of data because
it misses the rise in multiplicity due to mini-jets.
The predictions of both HIJING and HIJING/B\=B models are 
in better agreement with the data when the effects of both quenching and 
shadowing are included.
Note that with default energy loss (dE/dx=2 GeV/fm)
and constant $p_0$=2 GeV/c, the energy
dependence obtained with HIJING is too rapid and the curves in 
Fig.~\ref{fig:fig2_qm02} are for a reduced effective energy loss 
dE/dx = 0.5 GeV/fm. One observes, however, that even this 
smaller energy loss still leads to a more rapid dependence on energy
than seen in the data.
The  RQMD v2.4 curves were obtained  using the {\em cascade mode} 
and taken into account its {\em rescattering} and {\em color ropes}
effects with their default parameters.

\begin{figure} [hbt!]
\vskip 0.5cm
\centering
\epsfig{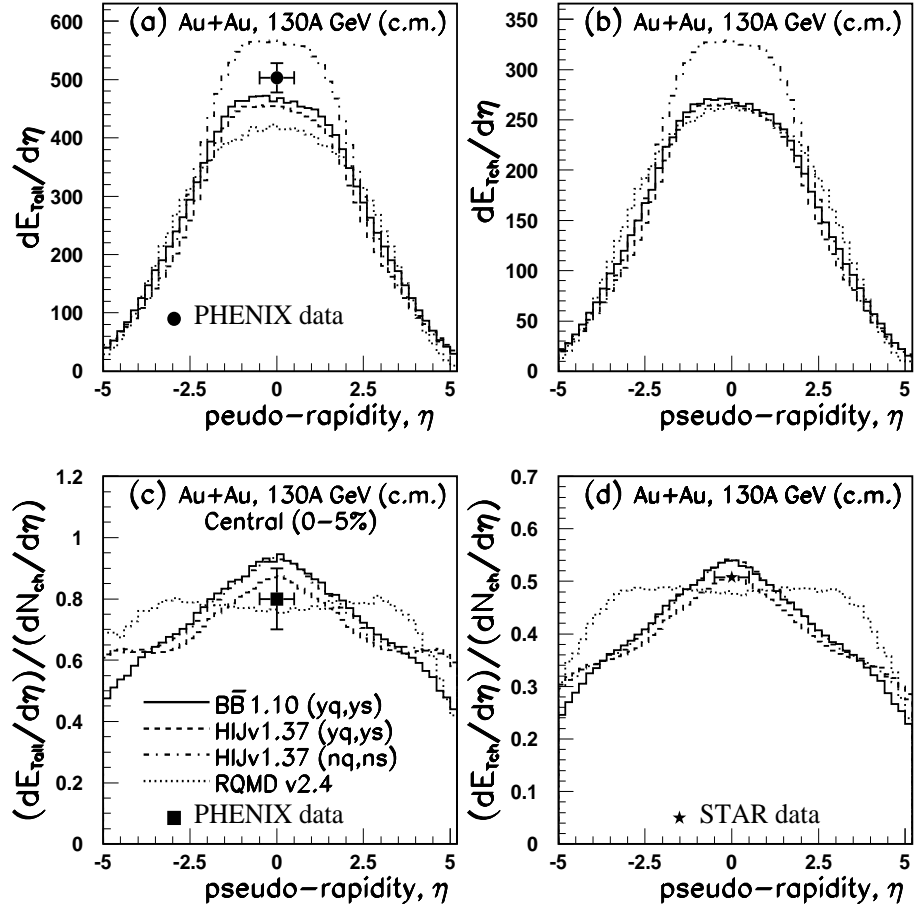} 
\vskip 0.5cm
 \caption[ch,et,et per charge,AA,RQMD]{\small 
Transverse energy distribution for: (a) all particles, (b) for 
charged particles only,
(c) total transverse energy per charged particles and 
(d) total transverse energy of charged particles per charged particles  
as a function of pseudo rapidity. Theoretical predictions 
 from HIJING/B \=B v1.10 {\em (yq,ys)}-solid,
HIJING v1.37 {\em (yq,ys)}-dashed, HIJING v1.37 {\em (nq,ns)}-dash dotted
and RQMD v2.4-dotted histograms are shown.
The data are from PHENIX \cite{phenix01_30} and STAR \cite{star_ad}.
The error bars for PHENIX include systematic uncertainties.}
\label{fig:fig3_qm02}
\end{figure}

Another global probe of the dynamics is the transverse energy per
charged particle, $dE_T/d\eta$ \cite{phenix01_20}.
This distribution is sensitive to $PdV$ work done by the plasma in
hydrodynamical models. In HIJING its value depends again on the
assumed shadowing and energy loss as seen 
in Fig.~\ref{fig:fig3_qm02}. In part (a) the total charged plus
neutral transverse energy distribution is shown. 
In part (b) the contribution from only charged particle is shown.
The results are for central (0-5 \%) Au+Au collisions 
at $\sqrt{s_{NN}}$=130 GeV.
Both versions HIJING (yq,ys) and HIJING/B\=B seem to account better for 
the observed $dE_T/d\eta\approx 500$ GeV than RQMD. This is again due
to the absence of mini-jets in RQMD model. 

We recall that FSS saturation model tend
in contrast to over-predict\cite{ekrt01} by a factor 2-3
the transverse energy because the saturation 
scale $p_s \sim 1$ GeV/c is significantly smaller
than the default $p_0=2$ GeV/c needed in HIJING to fit $p+p$ data.
Therefore, FSS requires reduction of the initial transverse energy 
due to longitudinal
hydrodynamic work. The same general tendency of over-predicting 
the transverse energy
is found in classical Yang Mills simulation of A+A \cite{iancu03}.
However, no detailed predictions of the transverse energy have been 
made within the KLN version\cite{dima_th01,dima_th02,dima_th03} of ISS models.

Another important difference between the predictions of models is
in the rapidity dependence of the transverse energy per particle
(see Fig.~\ref{fig:fig3_qm02}c and  ~\ref{fig:fig3_qm02}d). 
While RQMD predicts
a relatively constant value between $-2.5 \leq \eta \leq 2.5$,
both the numerator and denominator disagree with the data.
HIJING gives on the other hand a rather strongly 
peaked distribution at mid-rapidity.
This peaked distribution in HIJING is due to the localization of mini-jet
production to central rapidities.
Hydrodynamic models\cite{heinz03} generally assume a uniform 
boost invariant form of this ratio. 
We note that the PHOBOS observation
\cite{Back:2002gz} of a triangular dependence of the elliptic flow
$v_2(y)$ peaked at mid rapidity is very similar to the triangular pattern
of  $(dE_{Tall}/d\eta)/(dN_{ch}/d\eta)$ predicted by HIJING due to mini-jet
production at 130A GeV. We are not aware of any predictions
for this important global observable from saturation models.
Fig. 3d suggests that the initial conditions for hydrodynamics are not well
approximated by Bjorken boost invariant forms\cite{heinz03} assumed thus far.
 A full 3+1D hydrodynamical
simulation\cite{Hirano:2003pw}
 with such more realistic boost variant initial conditions should
be investigated to try to account for the PHOBOS elliptic flow.

The PHENIX data \cite{phenix01_20} show
a value closer to 0.8 GeV for $(dE_{Tall}/d\eta)/(dN_{ch}/d\eta)$
that is remarkably independent of $\sqrt{s_{NN}}\,\,$ from 17 GeV to 130 GeV 
and also independent of centrality. 
The observed independence on energy and centralities 
is very interesting since it is difficult to obtain 
such an effect in any transport theory 
with pQCD relaxation rates \cite{Molnar:2001ux,mik_th01}.

\begin{figure} [hbt!]
\vskip 0.5cm
\centering
\epsfig{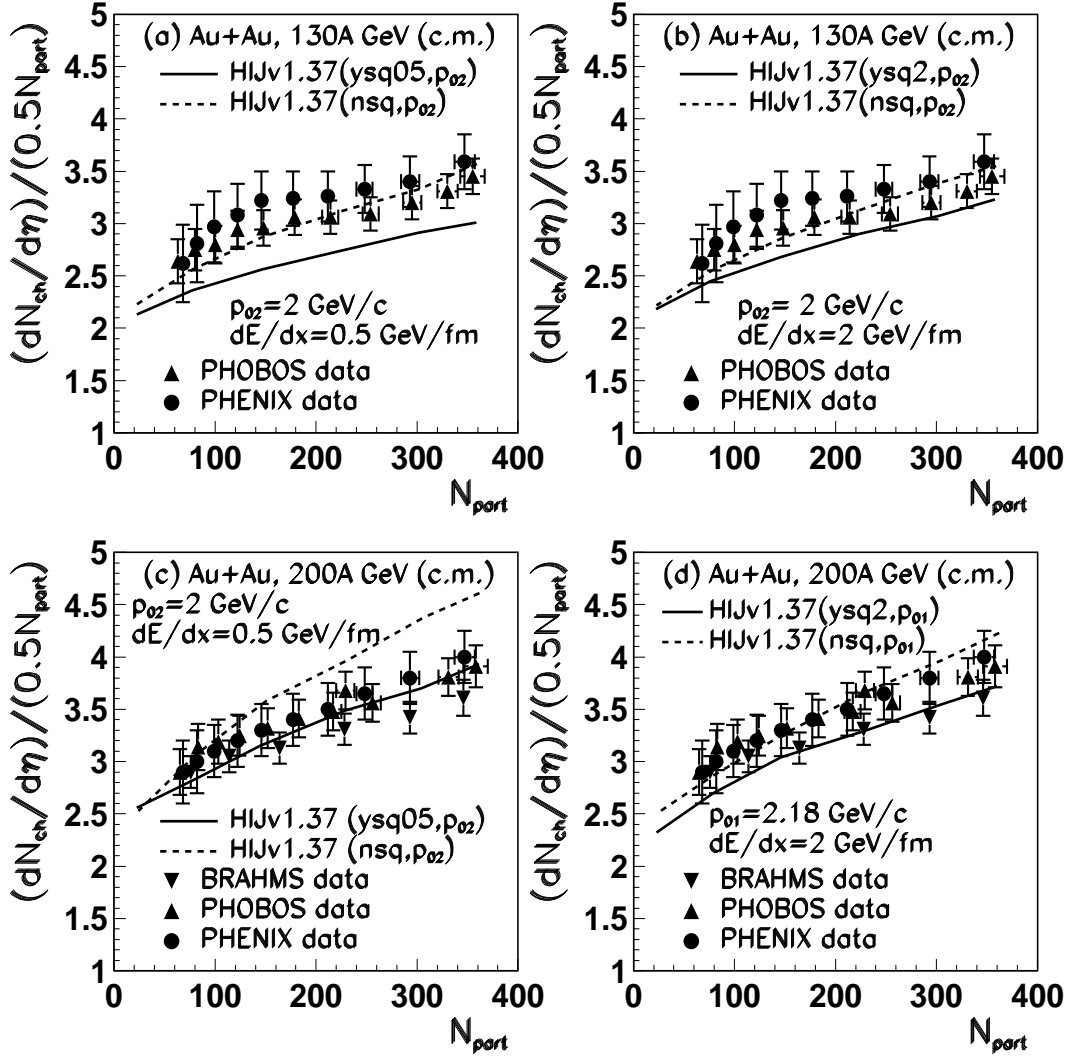} 
\vskip 0.5cm
\caption[dn_ch,dn_et as function of n_part]{\small
 Midrapidity $dN_{ch}/d\eta$ 
{\em per participant pair} as a function of the number of participants
at $\sqrt{s_{NN}}$=130 GeV (upper part).
 Theoretical predictions from  HIJING v1.37 model
with (ysq-solid lines) and without (nqs-dashed lines)
the effects of quenching and shadowing.
The data at $\sqrt{s_{NN}}$=130 GeV are from 
PHENIX \cite{phenix01_30}, \cite{bazilev03}, PHOBOS \cite{back01_30}.
Lower part are the results at $\sqrt{s_{NN}}$=200 GeV. 
The data are from BRAHMS \cite{brahms12_01}, PHOBOS \cite{back01_30}
and PHENIX \cite{bazilev03}.
The parameters within  HIJING calculations are given in the figure.
The error bars include both 
statistical and systematic uncertainties.}
\label{fig:fig4_qm02}
\end{figure}
\begin{figure} [hbt!]
\vskip 0.5cm
\centering
\epsfig{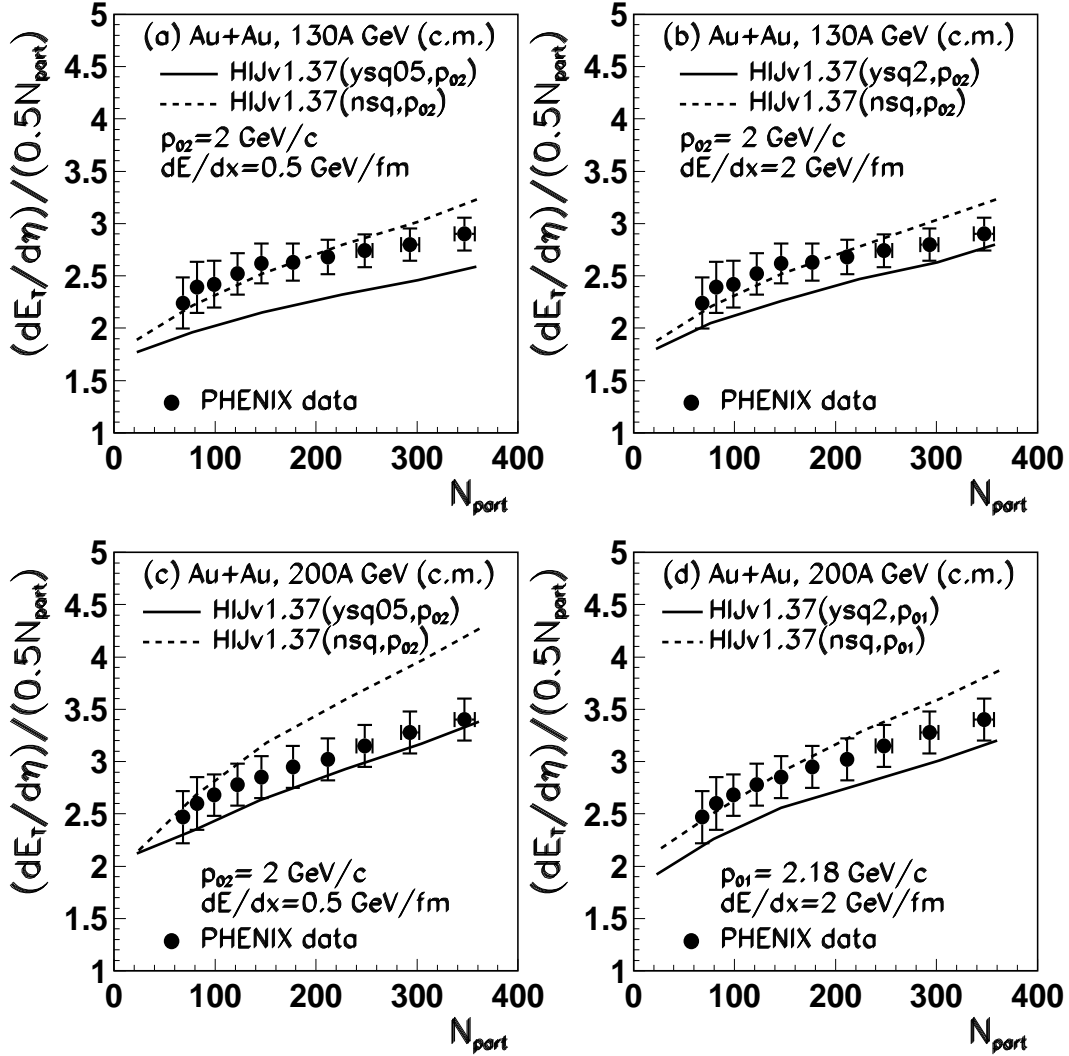} 
\vskip 0.5cm
 \caption[dn_ch,dn_et as function of n_part]{\small 
Midrapidity \protect{$dE_{T}/d\eta$} {\em per participant pair} as 
a function of the number of participants
at $\sqrt{s_{NN}}$=130 GeV (upper part) and $\sqrt{s_{NN}}$=200 GeV
(lower part). Theoretical predictions from  HIJING v1.37 model
with (ysq-solid lines) and without (nqs-dashed lines)
the effects of quenching and shadowing.
The solid and dashed lines have the same meaning as in Fig. 4.
The data at \protect{$\sqrt{s_{NN}}$}=130 GeV are from 
PHENIX \cite{phenix01_30},\cite{bazilev03}, PHOBOS\cite{back01_30}
and at $\sqrt{s_{NN}}$=200 GeV are from 
BRAHMS \cite{brahms12_01}, PHOBOS \cite{back01_30}
and PHENIX \cite{bazilev03}.
The error bars include both statistical and systematic uncertainties.}
\label{fig:fig5_qm02}
\end{figure}

We investigate next in more detail the centrality dependence
of $dN_{ch}/d\eta$ (Fig. 4) and  
$dE_{T}/d\eta$ (Fig. 5) per pair of participating nucleons.
Figure \ref{fig:fig4_qm02} and Fig. \ref{fig:fig5_qm02}
show the results for centrality dependence  
within HIJING v1.37 model calculations
at $\sqrt{s_{NN}}$ =130 GeV and $\sqrt{s_{NN}}$ =200 GeV
for {\em (yq,ys)} and {\em (nq,ns)}  scenarios 
in comparison with experimental data 
\cite{phenix01_30,bazilev03,back01_30,brahms12_01}. 
The parameters employed for these calculations are shown  on 
the figures.
We note that all HIJING curves extrapolate at low multiplicities 
to  the value $dN_{ch}/d\eta$=2.2, observed in p\=p collisions by the 
 UA5 collaboration \cite{ua5}.
The HIJING model predicts steady rise in the particle 
production per participant pair although the data seem to have 
a slower variation with $N_{part}$. The predicted increase is
due to nonlinear increase of 
 hard scatterings, which in contrast to the beam jet fragments,
dependent on the number of binary collisions. 
HIJING/B\=B predicts a similar trend although the calculated values
are lower than that given by HIJING by 10-15 \% and under-predicts 
the experimental results at $\sqrt{s_{NN}}$=130 GeV.

Figures 4a, 5a show the effect of lowering the energy loss to
0.5 GeV/fm as compared to a value of 2 GeV/fm used in Figs. 4b,5b. 
The default parameter
predictions at  $\sqrt{s_{NN}}$=130 GeV (Figs. 4b,5b) 
are more consistent with data.
However, for energy loss dE/dx =2.0 GeV/fm and $p_0$= 2.0 GeV/c 
assumed to be independent of $\sqrt{s}$ it is found
that the ratio of $R_{200/130}$ for midrapidity $dN_{ch}/d\eta$ 
is over-predicted for most central collisions
by $30\%$ as shown in ref. \cite{bazilev03}.
This is a major failing of the HIJING
assumption of energy independent mini-jet scale $p_0=2$ GeV/c.
Motivated by the energy dependence predicted by the saturation scales
in FSS and ISS models discussed in the introduction,
and the data, we study in Figs. 4d and 5d the effect of allowing
a slight increase with energy  from
$p_0(\sqrt{s})$=2.0 GeV/c at  $\sqrt{s_{NN}}$=130 GeV to
$p_{0}(\sqrt{s})$=2.18 GeV/c at $\sqrt{s_{NN}}$=200 GeV. 
Such an energy dependence was also found necessary in \cite{xin_li02} 
using more modern structure functions than in HIJING.

\begin{figure} [hbt!]
\vskip 0.5cm
\centering
\epsfig{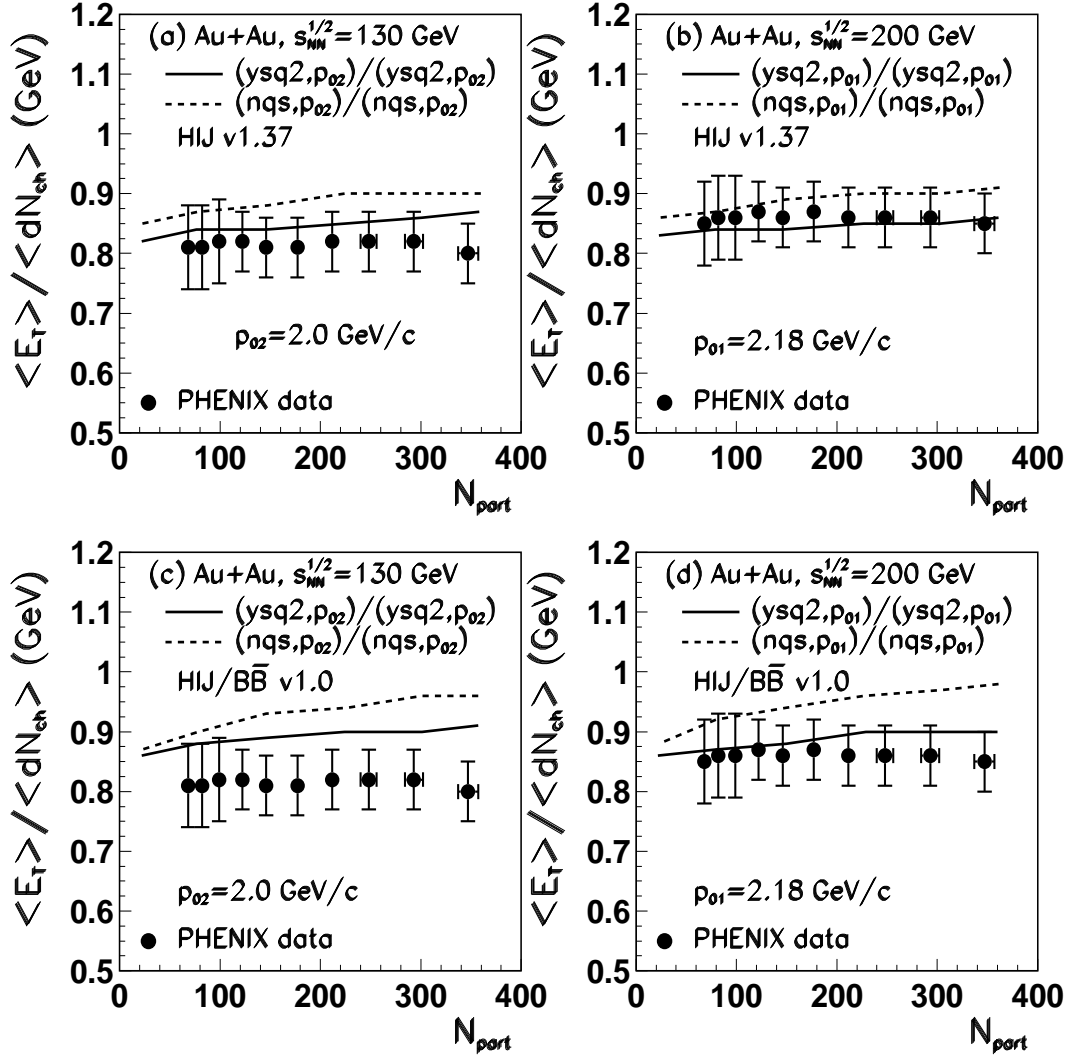} 
\vskip 0.5cm
 \caption[dn_ch,dn_et as function of n_part]{\small 
Midrapidity ratios 
(\protect({$dE_{T}/d\eta$})/{$dN_{ch}/d\eta$}) 
as a function of the number of participants
at $\sqrt{s_{NN}}$=130 GeV and  $\sqrt{s_{NN}}$=200 GeV 
Theoretical predictions within HIJING v1.37
(upper part) and HIJING/B\=B v1.10 (lower part).
The solid and dashed lines have the same meaning as in Fig. 4.
The data are from PHENIX\cite{bazilev03}.
The error bars include both statistical and systematic uncertainties.}
\label{fig:fig6_qm02}
\end{figure}

\begin{figure} [hbt!]
\vskip 0.5cm
\centering
\epsfig{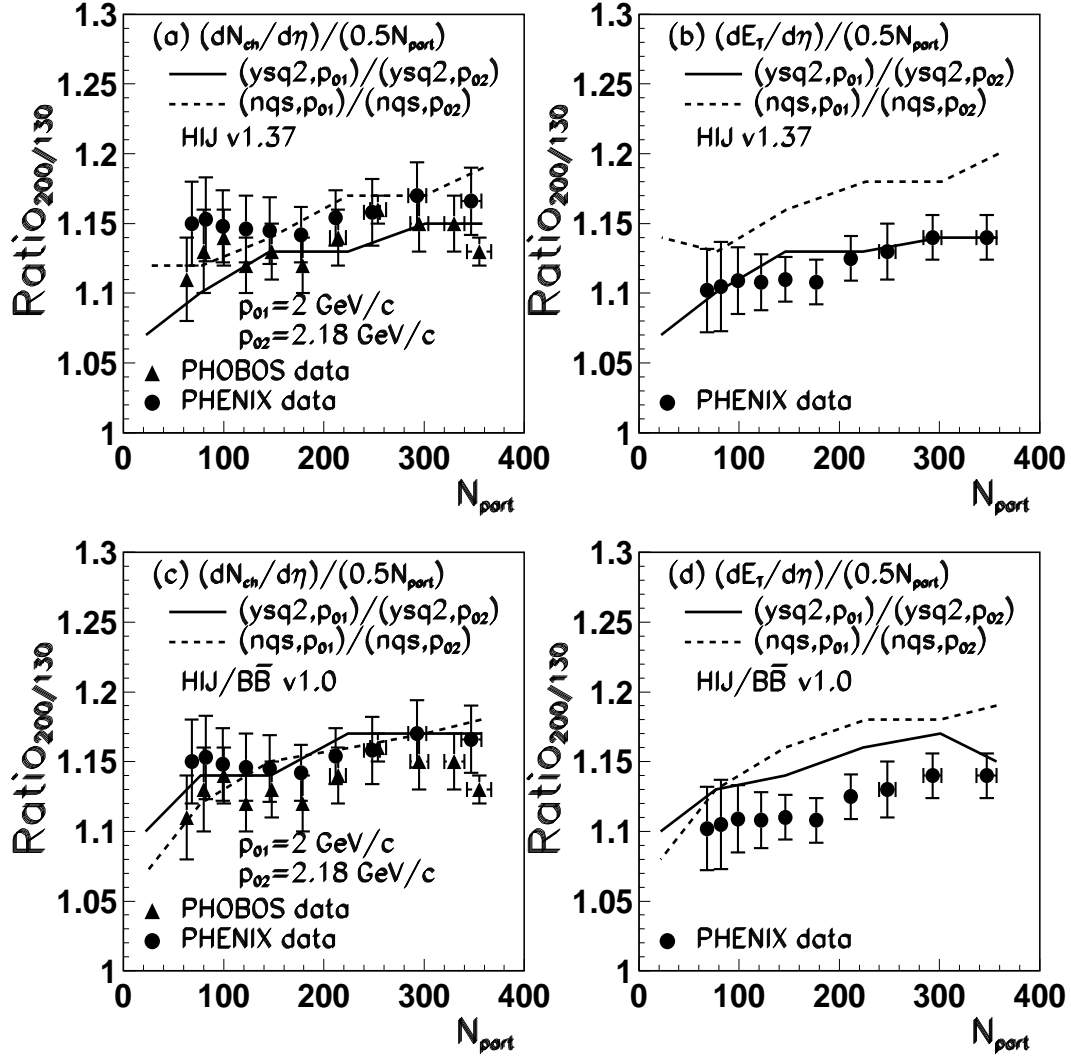}
\vskip 0.5cm
 \caption[dn_ch,dn_et as function of n_part]{\small
 Ratios $R_{200/130}$ for midrapidity $dN_{ch}/d\eta$ and $dE_{T}/d\eta$
{\em per participant pair} as a function of the number of participants.
Theoretical predictions within HIJING v1.37 (upper part) and   
HIJING/B\=B v1.10 (lower part) model 
with (ysq,$p_0$) and without (nsq,$p_0$)
the effects of quenching and shadowing are compared to  
data \cite{bazilev03}. For the labels of solid and dashed lines 
see Fig. 4(b),(d) and Fig. 5(b),(d).}
\label{fig:fig7_qm02}
\end{figure}

Figure~\ref{fig:fig6_qm02} ($E_{T}/N_{ch}$ transverse energy per
charged particle) and 
Fig.~\ref{fig:fig7_qm02} (ratios $R_{200/130}$ for 
midrapidity $dN_{ch}/d\eta$ and $dE_{T}/d\eta$) 
present the results obtained within both
models HIJING v1.37 (upper part) and HIJING/B\=B v1.10 (lower part). 
The data are much closer to the quench and shadowing 
{\em (yq,ys)} scenario.

It was shown \cite{brahms12_01} that hard scattering 
component to the charged particle production remains 
almost constant (36 $\pm$ 6) \% over the energy range 
$\sqrt{s_{NN}}$=130 GeV - 200 GeV. 
We see when comparing the values from Figs. 4d,5d to Figs. 4c,5c 
and especially from Fig. 6 and Fig. 7, that the energy dependence
of this global measure is reduced considerably 
to within the experiment range
by allowing a modest increase of $p_0$ without assuming any 
additional $N_{part}$ dependence of this scale.
From these results we conclude that a $10\%$ increase 
with energy of the mini-jet scale ($p_0$)
is required in both HIJING models to account for the
centrality and energy dependence of the global multiplicity and 
transverse energy observables.
 
\section{Jet Quenching and the Nuclear Modification Factor}

High $p_{T}$ hadron spectra
have been widely analyzed at SPS and RHIC energies \cite{miklos03}.
We  investigate in this section how well 
can  HIJING and HIJING/B\=B 
describe the high $p_T$ hadronic spectra in pp 
collisions and their predicted nuclear modifications in AA collisions.

The observation\cite{ad03_pi0_1,jia03_1,star01_raa,jet_star03}
of strong suppression of high 
$p_T$ hadron spectra in central $Au+Au$ at RHIC energies 
\cite{gyu90,wang92,baier95} is the most dramatic new
dynamical phenomena discovered at RHIC relative to SPS.
We recall that the comparison of parton
model calculations and the experimental data does not show 
any evidence of parton energy loss at SPS energies
\cite{Wang:1998hs}. 
The observed absence of quenching in $d+Au$,
\cite{jia03_1,star_dAu,phobos_dAu,brahms_03} 
as predicted in
Refs.\cite{wang92,Vitev:2002pf,Vitev:2003xu,Wang:2002ri,Levai:2003at},
proves that quenching is caused by final state interactions 
in the dense matter formed in Au+Au collisions and not due to gluon shadowing.

Parton-parton {\em hard scattering} with large momentum transfer 
produces  high momentum quarks or gluons 
which fragment into jets of hadrons.
The leading particles manifest themselves in a power-law like shape of the 
momentum distribution. High momentum partons 
are predicted to lose a significant fraction of their energy by 
gluon bremsstrahlung leading to a suppression 
of the high momentum tail 
of the single hadron inclusive spectra\cite{wang92}.
It has been argued that data from RHIC experiments show characteristic 
features consistent with such {\em ``jet quenching effects''}
\cite{phenix01_40,qm2001_drees,hirsh02}.
Other interpretations have been proposed after the data
became available. These are based on 
gluon saturation in the initial nuclear wave-function
\cite{dima03}, coherent fields and their geometry \cite{shuryak03}, 
surface emission of the quenched jets \cite{muller02}, 
final state hadronic interactions \cite{greiner03} and 
quark coalescence \cite{lin02}. 

The default HIJING implementation of jet quenching 
uses a simplified algorithm most closely resembling
surface emission. The energy loss is implemented by 
testing the number of interactions that a jet will have
along its propagation line with excited participant {\em strings}.
The approximate linear participant number scaling of the bulk multiplicity
motivates this approximation to the transverse matter density profile
through which the jets propagate.
The distance between collisions is fixed by a mean free path parameter,
$\lambda=1$ fm by default. Energy loss is implemented by splitting
the energy of the jet among multiple gluons 
with energies $\omega_i= \Delta z_i dE/dx$
where $dE/dx=2$(1) GeV/fm for gluon (quark) jets and $\Delta z_i$ are 
distances between collisions. This simplified mechanism  suppresses jets
that originate more than one mean free path from the surface.

The effect of the nuclear modification is 
quantified  in terms of the ratio\cite{xin_th03}
\begin{equation}
R_{AA}(p_{\perp})=\frac{d^{2}N_{AA}/dydp_{\perp}}
{<N_{coll}>d^{2}N_{pp}/dydp_{\perp}},
\end{equation}
where $<N_{coll}>$ is the average number of binary collisions of the
event sample and can be calculated from the nuclear overlap integral
($T_{AA}$) and the inelastic nucleon-nucleon cross section: 
$<N_{coll}>= \sigma^{inel}_{nn}<T_{AA}>$.
In the HIJING effective surface emission model, we can expect
$R_{AA}\propto (\lambda/N_{part}^{1/3}\;\rm{fm})$.

\begin{figure} [hbt!]
\vskip 0.5cm
\centering
\epsfig{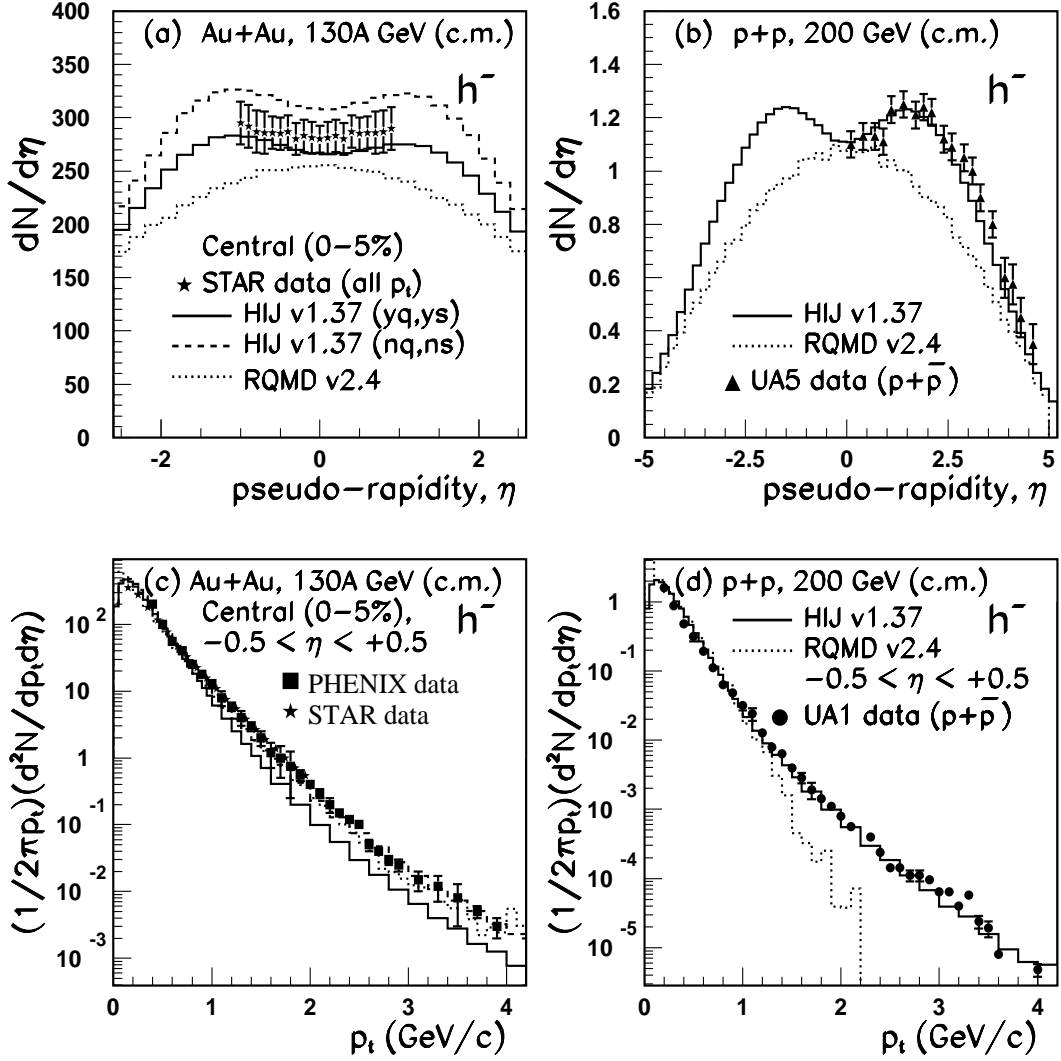} 
\vskip 0.5cm
 \caption[ch-_mik1_pp_cp_1]{\small
Comparison of Au+Au and p+p at RHIC energies is shown.
(a): Negative hadrons pseudo-rapidity 
 spectra for Au+Au collisions at $\sqrt{s_{NN}}$=130 GeV
including STAR central (0-5\%) data\cite{star_ad}.
(b) p+p at $\sqrt{s_{NN}}$=200 GeV including UA1 p+\=p
data\cite{ua5}.
HIJING v1.37 {\em (yq,ys)}-solid, HIJING v1.37 {\em (nq,ns)}-dashed, 
and RQMD v2.4-dot dashed histograms are theoretical calculations.
Parts c) and d) compare the $p_\perp$ distributions. 
PHENIX data \cite{phenix01_40}, STAR data \cite{star_ad} 
and UA1 data\cite{ua1} are shown. The error bars 
include systematic uncertainties in part (a) and are statistical 
only in parts (b,c,d). 
\label{fig:fig8_qm02}}
\end{figure}

The absolutely normalized transverse momentum spectra  
and pseudo rapidity distributions for Au+Au central (0-5\%) collisions
 \cite{phenix01_40,star_ad} 
at $\sqrt{s_{NN}}$ = 130 GeV are shown in
Fig.~\ref{fig:fig8_qm02}. We compare the STAR
\cite{star_ad}  pseudo-rapidity distribution 
of negative hadrons ($h^-$) for the central (0-5\%) 
Au+Au collisions with the 
predictions from HIJING {\em (yq,ys)}-solid line, 
 HIJING {\em (nq,ns)}-dashed line and RQMD (dash-dotted line).
The negative hadron pseudo rapidity distributions in part (a) are best 
reproduced with shadowing 
and quenching effects present. Note in Fig. 8c, however,  that the 
moderate $1<p_T<4$ GeV spectra are too strongly 
over-quenched by the default HIJING parameters. The ``no shadow,
no quench'' calculation  over-predicts the global rapidity 
density but remarkably fits the moderate $p_T$ spectrum rather well. 
So where is the energy loss? As we discuss below, 
it is likely camouflaged by anomalous  baryon excess.

Note that RQMD fits the moderate $p_T$ 
$h^-$ data better than HIJING (yq,ys).
However, as shown in Fig.~\ref{fig:fig8_qm02}b,
the absence of mini-jet production
in RQMD causes it to miss the observed features of the $pp$ 
rapidity density\cite{ua5} at $\sqrt{s_{NN}}$= 200 GeV 
that HIJING reproduces. Similarly, because of multiple jet production,
HIJING reproduces the power law like tail of the p+p $p_T$ spectra
very well while RQMD shows a glaring discrepancy (Fig. 8d).
This result is significant because it demonstrates that
the agreement of RQMD in Fig.8c is fortuitously due to 
a  strong nuclear dependent
Cronin like  multiple collisions algorithm enhancement.
This is similar to the fortuitous agreement\cite{Gyulassy:1998nc}
 of HIJING with WA98 moderate $p_T$
pion data at the SPS, where the results were shown 
to be exponentially sensitive to the Cronin algorithm adopted.

To better understand why HIJING over-predicts the quenching of $h^-$ for
$p_T<4$ GeV/c, we turn next to the latest data on identified
$\pi^0$ central (0-10 \%) Au+Au collisions at $\sqrt{s_{NN}}$= 200
GeV. We use these data because of much 
higher $p_T$ reach than at $\sqrt{s_{NN}}$=130 GeV.
Also in this case, $Au+Au$ can be compared directly to the new
p+p data measured by PHENIX\cite{ad03_pi0_1,ad03_pi0_2}.  
In Fig. 9a, the predicted $\pi^0$ $p_T$ spectra based on
$dE/dx$=0, 0.5, and 2 GeV/fm are compared to the Au+Au data out to 8 GeV/c.
In Fig. 9b the recent
$p+p\rightarrow \pi^0+X$ data are compared to default HIJING. This
shows  even more clearly than the original test\cite{hij92_99},
how well HIJING is able to reproduce the high $p_T$ spectrum
out to 10 GeV/c in the elementary $p+p$ case.
  The $R_{AA}(p_T)$ nuclear modification factor is shown in Fig. 9c.  
It is observed that the default energy loss parameters(dE/dx=2 GeV/fm)
describe very well the jet
quenching pattern of neutral pions in central collisions.
 The Lund string fragmentation
mechanism of hadronization in HIJING leads to a rather slow
growth of $R_{AA}$ to unity at high $p_T$
(dashed histogram in Fig. 9c). Only after $p_T>4$ GeV/c do the
details of hadronization become irrelevant, and in that range, the
default energy loss leads to a factor of ~five suppression, in
agreement with the naive surface emission estimate with $\lambda=1$
fm.  The relative suppression ratio 
$R_2 \equiv R_{AA}(ysq)/R_{AA}(nsq)$ (Fig. 9d) shows in more detail 
the leveling off of $R_2$ beyond 4 GeV/c for the default energy loss, 
while it levels off at 0.4 for a
reduced energy loss of 0.5 GeV/fm. 
Note that in all calculations the default
$p_0=2$ GeV/c was used, as it has, however, no effect at high $p_T$.

In Fig. 10 we compare the quenching pattern of HIJING to data for peripheral
(60-80\%) reactions. Fig. 10b is the $pp$ data scaled by 
the number of collisions in this peripheral reaction class. 
The main difference with respect to Fig. 9 is that unlike 
in central collisions, even a reduced energy 
loss $dE/dx=0.5$ GeV/fm over-predicts the small modification
of $R_{AA}$ from unity observed for more peripheral interactions. 
As more clearly seen in Fig. 10d,
HIJING predicts a 30\% suppression in peripheral 
interactions that is not seen in the data.
We conclude that while the central reaction suppression 
is correctly predicted by HIJING,
the surface emission algorithm adopted to model energy 
loss is not realistic
and does not reproduce the centrality dependence observed by PHENIX.

We have studied the dependence of the quenching pattern on variations of
the mean free path parameter ($\lambda$) of HIJING as well. We find that the
quenching is indeed sensitive to $\lambda$. In central collision 
increasing $\lambda$ to 3 fm for example decreases the quenching for
$dE/dx=2$ GeV/fm by a factor of approximately two. To explain the
 centrality dependence an $N_{part}$ dependence of $\lambda$ must be assumed.
We do not pursue here such an elaboration of the HIJING energy loss algorithm
but future  studies would be desirable to 
help distinguish between surface emission 
and volume emission models of jet energy loss.

The sensitivity of the nuclear modification factor $R_{AA}(p_{\perp})$
  to the mean energy loss parameter in  HIJING  at the lower 130A GeV energy
is more clearly revealed in the high statistics computation shown
in Fig.~\ref{fig:fig11_qm02}. The data at 
$\sqrt{s_{NN}}$= 130 GeV are taken from PHENIX
\cite{phenix01_40} and STAR \cite{star01_raa}.
This figure extends the comparison in Fig. 8 to the range $p_T\sim7$ GeV/c.
The main discrepancy between the  negative hadrons $h^-$ data and  calculations
is the observed distinct localized bump with 
a maximum at $p_T\sim 2$ GeV/c.
$R_{AA}$ approaches the predicted quenching pattern 
with $dE/dx=0.5$ GeV/fm  only at the highest $p_T$ measured. 

PHENIX has found  that the excess negative hadrons in the 2-4 GeV range
are in fact due to anti-protons\cite{qm2001_za}.
In order to check whether an enhanced baryon 
junction loop\cite{svance99} mechanism 
could possibly account for that  
excess, we plot $R_{AA}$ for HIJING/B\=B\cite{svance_cu} in Fig. 12.
While the junction source reduces the discrepancy 
between data and HIJING shown in Fig. 12a, 
it fails to account for the very large excess of anti-protons at 
moderate $p_T$. From Fig. 12a,b, we see that 
junctions as currently implemented
do not solve the centrality dependence
problem discussed in connection with Fig. 10 either.

As a final comment we compare the ``central to peripheral'' 
nuclear modification factor  defined by
\begin{equation}
R^{cp}_{AA}(p_{\perp})=\frac{(Yield/<N_{coll}>)_{(0-5\%)}}
{(Yield/<N_{coll}>)_{(60-80\%)}}
\end{equation}
where $Yield$ =
($1/N_{events}$)($1/2\pi\,p_{\perp}$)($d^2N/dp_{\perp}d\eta$)
to calculated values for this particular ratio.
The data are from PHENIX \cite{phenix01_40,phe02_raa_1}
and STAR  \cite{star01_raa}.
Even though B\=B version fails to describe both 
the numerator (Fig. 12a) and the denominator
(Fig. 12b), it accidentally describes the $R^{cp}_{AA}$ ratio 
of central collisions (Fig. 13a).
No such lucky coincidence occurs for 
peripheral collisions using the default HIJING (Fig. 13b).
This figure  demonstrates the great care one must exercise 
in interpreting any agreement of dynamical models
with specific ratios. It is always essential to check whether
the model is able to reproduce
the absolutely normalized  spectra, as in Figs. 9 and 10. Only 
after a model passes that test can  any 
agreement with specific data ratios be considered seriously.

\section{Summary and Conclusions}

We have investigated in this paper how the predictions of HIJING and RQMD
exclusive nuclear collision event generators
compare to the new available data from RHIC. 
We concentrated on two classes of observables.
First the global number and transverse energy distribution in rapidity
was considered. Then we focused on the new jet quenching nuclear modification
factors. 

The energy dependence of global observables rule out RQMD because
of its neglect of hard  pQCD mini-jet production.
However, the observed  energy dependence also rules out HIJING 
in its default parameter settings. 
The separation scale, $p_0$ between soft and hard processes, which
assumed in HIJING to be a constant 2 GeV/c independent
of $\sqrt{s}$ and centrality,  predicts a too rapid growth of multiplicity.
Motivated by FSS and ISS parton saturation models and the data,
we tested and found that allowing 
a 10\% growth of $p_0$ from 2.0 to 2.18 GeV greatly improved
the consistency of HIJING results  with the observed RHIC systematics.
In all cases, the default shadowing (with identical
quark and gluon shadowing) assumed was found to be
essential to reduce the mini-jet flux and not to over-predict
the multiplicity. The small enhancement of the multiplicity
due to jet quenching with the default energy 
loss was consistent with experimental data
once the small energy dependence of $p_0$ is taken into account.

Our analysis of the jet quenching pattern predicted by HIJING
shows that the default $dE/dx=2$ GeV/fm accounts remarkably well
for the suppression pattern of $\pi^0$ out to $p_T=8$ GeV/c 
as observed for central $Au+Au$ 
collisions at $\sqrt{s_{NN}}$=200 GeV. 
A major advantage of HIJING over other models
is that it reproduces accurately both the low $p_T$
dominated rapidity density and the high $p_T$ recent 
200 GeV  $p+p\rightarrow \pi^0+X$ data\cite{ad03_pi0_2} at the same time. 

However, neither HIJING nor HIJING/B\=B are able to account for the 
anomalous baryon lump
in the intermediate $p_T<4$ GeV/c region.
Furthermore, we noted that the energy loss algorithm in HIJING corresponds
effectively to surface emission with a  
default $\lambda=1$ fm mean free path.
We checked that increasing $\lambda$ leads to less suppression.
We found that a constant $N_{part}$ independent
$\lambda$, however, is not compatible
with the observed centrality dependence of jet quenching.

The failure of the current implementation of baryon junction loops
in HIJING/B\=Bv1.10 to reproduce the observed $p_T$ enhancement
of anti-baryons and baryons needs, however, 
further study since an enhancement
was theoretically anticipated\cite{kharzeev96,svance99}. 
We are currently investigating
why this feature of baryon junction dynamics did not 
emerge from numerical simulations
with this code. Understanding the physical origin of the 
(anti) baryon anomalies
is essential to disentangle competing mechanisms such as 
collective hydrodynamic flow\cite{heinz03},  
multi-quark coalescence\cite{lin02},
and possibly novel baryon junction dynamics\cite{Vitev:2001zn} at RHIC.


\section{Acknowledgments}

{\bf Acknowledgments:} The authors would like to thank Subal Das Gupta
for careful reading of the manuscript and to Stephen 
Vance for useful discussions.
This work was partly supported by the Natural Science and 
Engineering Research Council of Canada and the 
``{\it Fonds Nature et Technologies}'' of Quebec.  
This work was supported also by the Director,
Office of Energy Research, Office of High Energy and Nuclear Physics,
Division of Nuclear Physics, and by the Office of Basic Energy
Science, Division of Nuclear Science, of the U. S. Department 
of Energy under Contract No. DE-AC03-76SF00098 and
DE-FG-02-93ER-40764.


\newpage 

\newpage 

\newpage

\newpage

\newpage

\newpage 

\newpage

\newpage 

\begin{figure} [hbt!]
\vskip 0.5cm
\centering
\epsfig{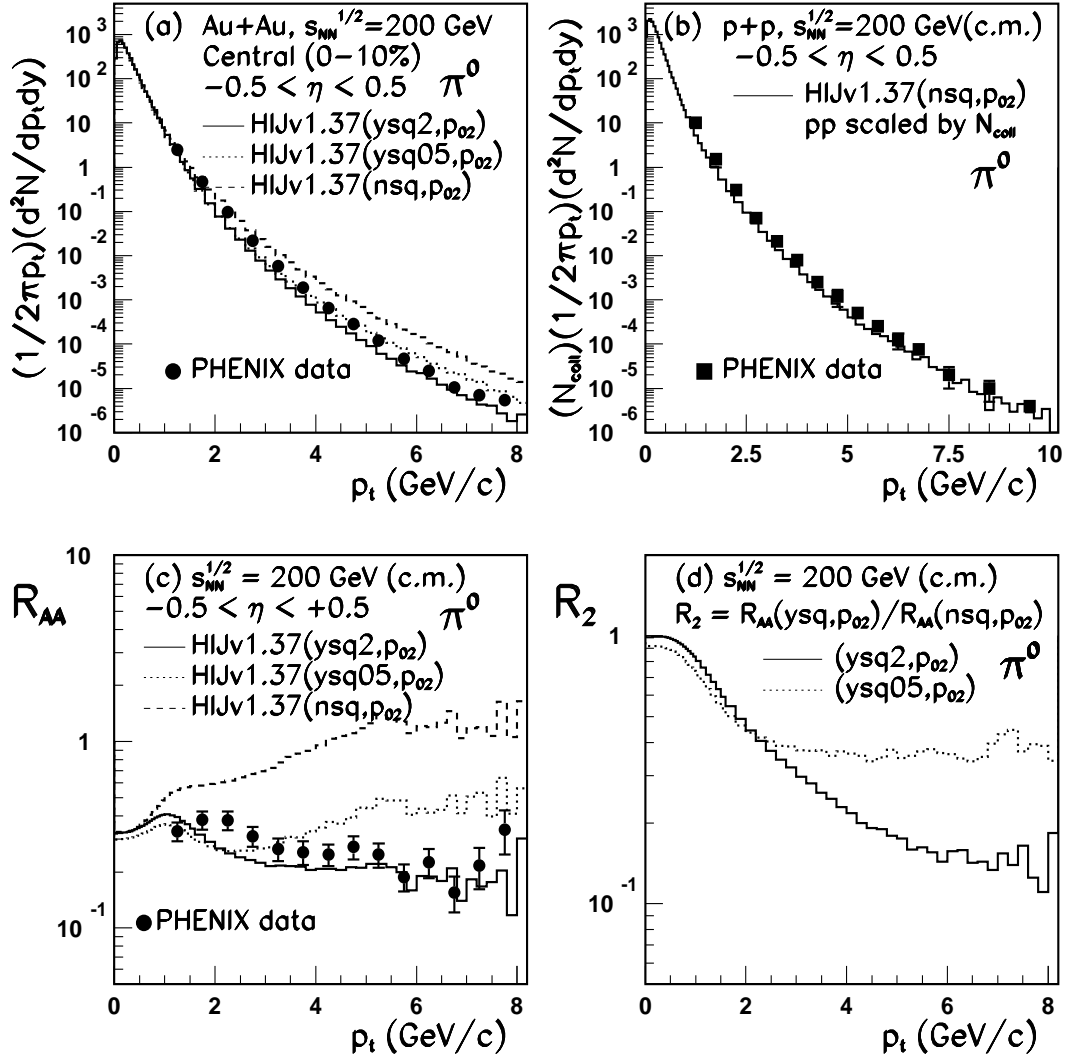} 
\vskip 0.5cm
\caption[R_AA vs p_t AApp for pi0]{\small
Invariant yields at midrapidity for $\pi^0$ in 
central (0-10 \%) Au+Au collisions (Part a)
and scaled by number of binary collisions ($N_{coll}$) 
in p+p interactions at $\sqrt{s_{NN}}$=200 GeV.
Nuclear modification factor $R_{AA}$ (Part c)
and ratio $R_2$ (Part d)
as a function of transverse momentum as predicted by HIJING v1.37
with  (solid and dotted histogram) and
without (dashed histogram) shadowing and quenching effects.
The labels have the same meaning as in Fig. 4.
The data are from PHENIX \cite{jia03,ad03_pi0_1,ad03_pi0_2}. 
Only statistical error bars are shown.}
\label{fig:fig9_qm02}
\end{figure}

\newpage 

\begin{figure} [hbt!]
\vskip 0.5cm
\centering
\epsfig{file=fig10_qm02.epsi,height=14cm, width=14cm} 
\vskip 0.5cm
\caption[Raa vs pt AA]{\small
Invariant yields at midrapidity for $\pi^0$ in 
central (60-80 \%) Au+Au collisions (Part a)
and scaled by number of binary collisions ($N_{coll}$) 
in p+p interactions at $\sqrt{s_{NN}}$=200 GeV.
Nuclear modification factor $R_{AA}$ (Part c)
and ratio $R_2$ (Part d)
as a function of transverse momentum as predicted by HIJING v1.37
with  (solid histogram) and
without (dashed histogram) shadowing and quenching effects.
The labels have the same meaning as in Fig. 4.
The data are from PHENIX  \cite{jia03,ad03_pi0_1,ad03_pi0_2}. 
Only statistical error bars are shown.}
\label{fig:fig10_qm02}
\end{figure}

\newpage

\begin{figure} [hbt!]
\vskip 0.5cm
\centering

\epsfig{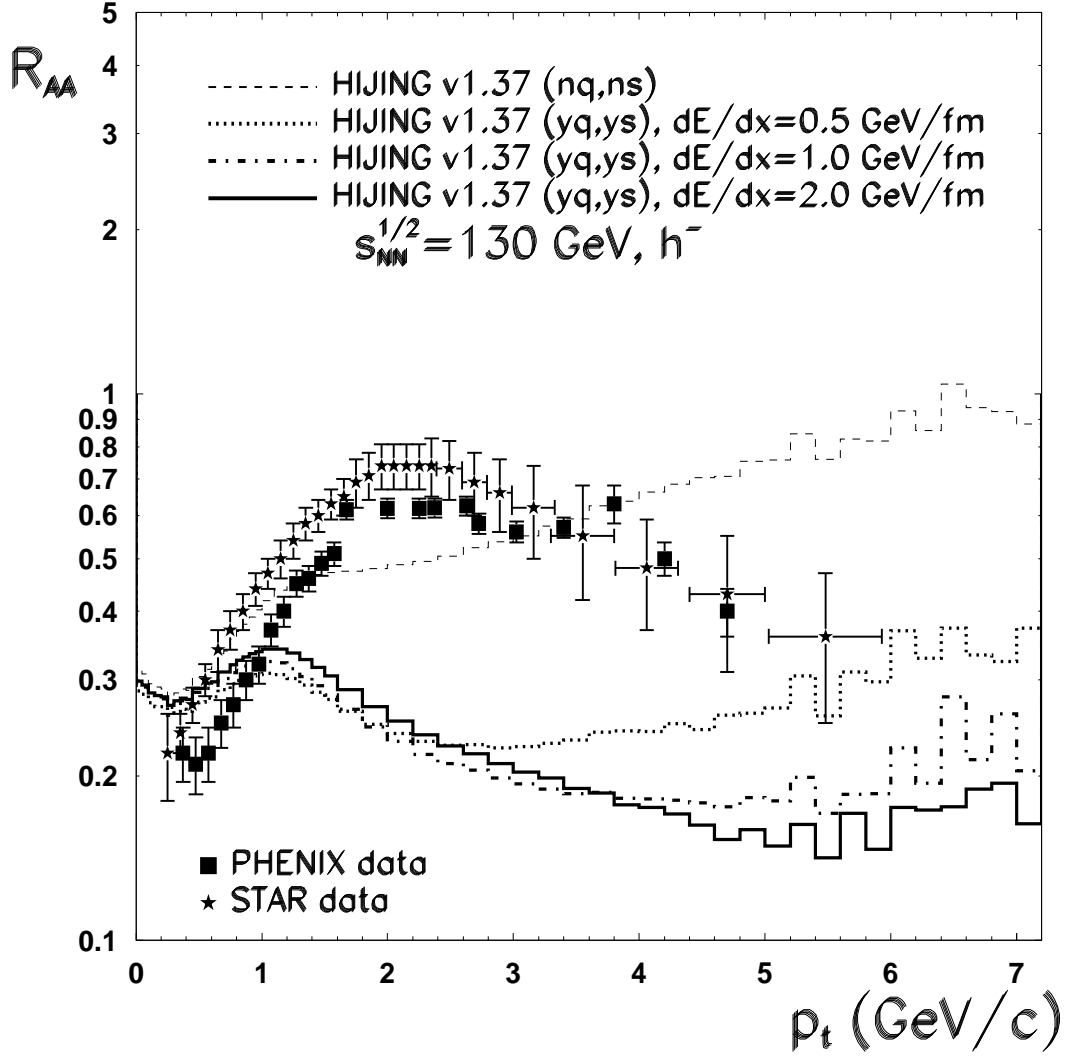} 
\vskip 0.5cm
\caption[R_AA vs p_t AApp]{\small
Nuclear modification factor $R_{AA}$,
as a function of transverse momentum as predicted by HIJING v1.37
for central (0-5\%) Au+Au collisions at $\sqrt{s_{NN}}$=130 GeV  . 
The different histograms are the results obtained without quenching 
and with quenching assuming different values for parton energy loss.
The data are from PHENIX  \cite{phenix01_40} (squares) and from 
STAR \cite{star01_raa} (stars). Only statistical error bars are shown.}

\label{fig:fig11_qm02}
\end{figure}

\newpage 

\begin{figure} [hbt!]
\vskip 0.5cm
\centering

\epsfig{file=fig12_qm02.epsi,height=14cm, width=14cm} 
\vskip 0.5cm
\caption[Raa vs pt AA]{\small
Comparison of the nuclear modification factor ($R_{AA}$)
for Au+Au as predicted by HIJING/B\=B v1.10 (upper part)
and HIJING v1.37 (lower part) for central (0-5\%)
and peripheral (60-80\%) Au+Au collisions.  
The data are from PHENIX
\cite{phenix01_40,phe02_raa_1} and STAR \cite{star01_raa}. 
Only statistical error bars are shown.}
\label{fig:fig12_qm02}
\end{figure}

\newpage 

\begin{figure} [hbt!]
\vskip 0.5cm
\centering

\epsfig{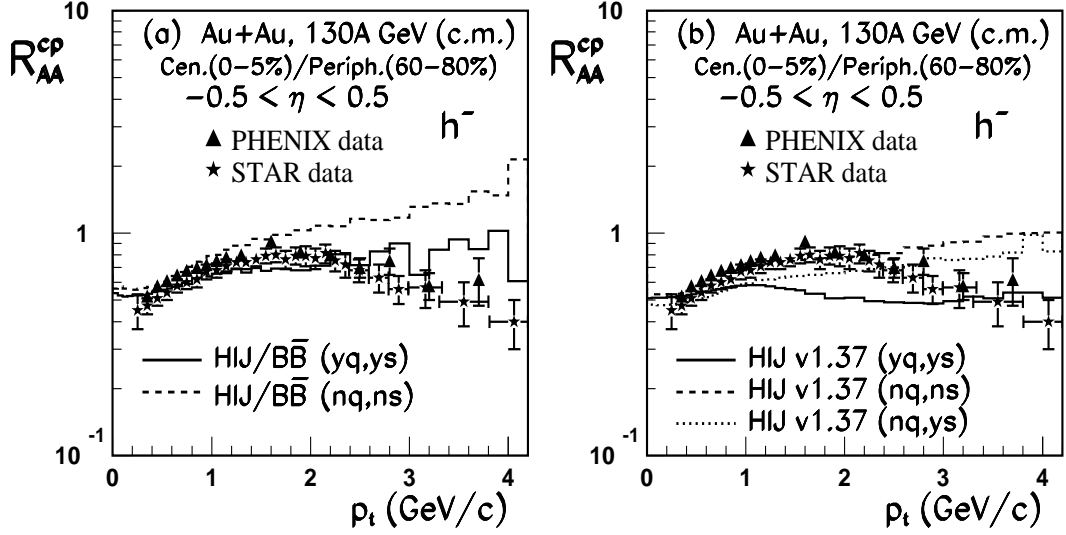} 
\vskip 0.5cm
\caption[Raa_cp vs pt AA 5\%/6080\%]{\small
Comparison of the nuclear modification factor ($R_{AA}^{cp}$)
for Au+Au as predicted by HIJING/B\=B v1.10 (part (a))
and HIJING v1.37 (part (b)) for ratio of central (0-5\%)
to peripheral (60-80\%) collisions.  
The data are from PHENIX
\cite{phenix01_40,phe02_raa_1} and STAR \cite{star01_raa}. 
Only statistical error bars are shown.}
\label{fig:fig13_qm02}
\end{figure}

\end{document}